\documentclass[runningheads]{llncs}
\usepackage{graphicx}
\usepackage{url}
\usepackage{csquotes}
\usepackage{float}
\MakeOuterQuote{"}

\begin{document}
\title{Human-Artificial Interaction in the Age of Agentic AI: A System-Theoretical Approach}
\titlerunning{Human-Artificial Interaction in the Age of Agentic AI}
\author{Uwe M. Borghoff \inst{1}\thanks{Corresponding author: uwe.borghoff@unibw.de}
\and Paolo Bottoni \inst{2} 
\and Remo Pareschi \inst{3}
}
\authorrunning{U. M. Borghoff, P. Bottoni \& R. Pareschi}
\institute{Computer Science, University of the Bundeswehr Munich, Neubiberg, Germany
\and
Computer Science, Sapienza University of Rome, Rome, Italy \\
\and
Stake Lab, Università degli Studi del Molise, Campobasso, Italy
}
\maketitle              % typeset the header of the contribution

\begin{abstract}
This paper presents a novel perspective on human-computer interaction (HCI), framing it as a dynamic interplay between human and computational agents within a networked system. 
Going beyond traditional interface-based approaches, we emphasize the importance of coordination and communication among heterogeneous agents with different capabilities, roles, and goals. 
A key distinction is made between multi-agent systems (MAS) and Centaurian systems, which represent two different paradigms of human-AI collaboration. MAS maintain agent autonomy, with structured protocols enabling cooperation, while Centaurian systems deeply integrate human and AI capabilities, creating unified decision-making entities.

To formalize these interactions, we introduce a framework for communication spaces, structured into surface, observation, and computation layers, ensuring seamless integration between MAS and Centaurian architectures, where colored Petri nets effectively represent structured Centaurian systems and high-level reconfigurable networks address the dynamic nature of MAS.

Our research has practical applications in autonomous robotics, human-in-the-loop decision making, and AI-driven cognitive architectures, and provides a foundation for next-generation hybrid intelligence systems that balance structured coordination with emergent behavior.
\end{abstract}

%%%%%%%%%%%%%%%%%%%%%%%%%%%%%%%%%%%%%%%
\section{Introduction}\label{sec:intro}
%%%%%%%%%%%%%%%%%%%%%%%%%%%%%%%%%%%%%%%
Agentic AI systems---capable of iterative planning, autonomous task decomposition, and continuous learning---are rapidly reshaping the landscape of human–computer interaction (HCI). 
Recent advances in Large Language Models (LLMs) and advanced conversational agents have revitalized the field of multi-agent systems, whose roots in Artificial Intelligence predate the current rise of generative AI. 
Historically, multi-agent systems relied on agents with relatively constrained capabilities; however, the emergence of powerful, conversationally adept LLMs significantly broadens the scope of possible multi-agent interactions. 
In this new paradigm, humans themselves can participate as fully capable agents, thanks to their innate conversational and decision-making faculties. The result is a complex, interconnected ecosystem \cite{Hopf2024}.

Simultaneously, novel and more direct forms of human–AI integration are emerging \cite{pareschi2024beyond,saghafian2024effective}, in which advanced AI cognition is harnessed to create ``Centaurian'' intelligence---a seamless fusion of human and machine-driven capabilities.
These developments push beyond traditional HCI in two distinct directions that mirror fundamental patterns found in biological systems \cite{vonBertalanffy1968,BorghoffBMP95,Miller1978}. Multi-agent systems, like biological ecosystems, emphasize how collections of autonomous entities can coordinate effectively while preserving distinct boundaries and roles. In contrast, ``Centaurian'' systems pursue deeper integration---analogous to symbiotic relationships in nature---fusing human and artificial competencies in tightly knit partnerships that often blur the lines between human decision-making and AI-driven processes. Living systems theory \cite{Maturana1980} helps us understand how both approaches must address a core challenge: maintaining system identity through regulated boundaries and feedback loops \cite{Wiener1948}, whether in loosely coupled collectives or tightly integrated hybrid intelligences.

Herbert Simon's influential work on cognitive architectures \cite{Simon1996} provides additional insight into how such human-AI integration might be structured. His tripartite model---comprising an external interface, a coding mechanism, and an internal processing system---was originally conceived to explain human problem-solving but extends naturally to hybrid systems. When human abilities merge with artificial intelligence, the resulting configuration can evolve from traditional tool use (\textit{Homo Faber}) to a more profound integration (\textit{Centaurus Faber}) where each component may be instantiated by either human cognition or AI subsystems \cite{pareschi2024beyond}. This perspective suggests how system boundaries might be organized to support both autonomous operation and deep integration.
To address these architectural challenges, we introduce a formal framework based on Petri nets that can model both multi-agent and Centaurian paradigms while supporting the key requirements suggested by living systems theory and Simon's architecture: clear boundaries, regulated interactions, and adaptive feedback loops. Our approach, grounded in rigorous process modeling, provides a unified way to specify how heterogeneous agents---human and artificial---can coordinate their activities while either maintaining autonomy or achieving deeper integration as needed.

The remainder of this paper is organized as follows. 
Section~\ref{sec:paradigms} analyzes two fundamental paradigms in human–machine collaboration---the multi-agent approach and the Centaurian one---in light of the recent resurgence of agentic AI. 
Section~\ref{sec:tech-background} provides foundational background on Petri nets and their extensions. 
Section~\ref{sec:comm-spaces} introduces the concept of communication spaces as a unifying framework for these paradigms, grounded in formal architectures and coordination mechanisms. 
Sections~\ref{sec:usecase1} and~\ref{sec:usecase2} illustrate these ideas with two use cases, demonstrating how communication spaces can be applied in practical HCI scenarios. 
Section~\ref{sec:relwork} discusses related work, and 
Section~\ref{sec:concls} concludes with implications and future directions.

\newpage

%%%%%%%%%%%%%%%%%%%%%%%%%%%%%%%%%%%%%%%%%%%%%%%%%%%%%%%%%%%%%%%%%
\section{Paradigms in Human-AI Integration}\label{sec:paradigms}
%%%%%%%%%%%%%%%%%%%%%%%%%%%%%%%%%%%%%%%%%%%%%%%%%%%%%%%%%%%%%%%%%
The emergence of agentic AI systems has given new life to two distinct approaches for human-AI collaboration: multi-agent architectures and Centaurian integration. 
This section examines these paradigms in detail, highlighting the distinct challenges each presents for system design and coordination.

Indeed, although both paradigms involve interactions between intelligent entities, they represent fundamentally different approaches to human-artificial collaboration \cite{pareschi2024beyond}. 
Multi-agent systems maintain distinct boundaries between components while enabling complex interactions, much like natural ecosystems. 
In contrast, Centaurian systems mirror the tight integration seen in biological organisms, where components merge functionally to create new capabilities \cite{saghafian2024effective}.

\subsection{Architectural Differences}
The primary architectural distinction lies in how these systems organize and maintain their identity \cite{pareschi2024beyond}. 

Multi-agent systems (MAS) emphasize functional independence, where agents ---human or artificial---operate as distinct entities with decision-making capabilities and dynamic collaboration. This design has allowed MAS implementations to thrive even with relatively simple AI agents, focusing on coordination rather than deep integration.

Centaurian systems, by contrast, create unified composite entities where human and artificial components become functionally interdependent. Theoretical foundations for such systems trace back to \cite{licklider1960man} and \cite{engelbart1962augmenting}, but their practical feasibility has emerged only recently with generative AI systems capable of complementing human cognitive functions.

A key concept in both systems is that of \textit{communication spaces}---regions of interaction where information exchange and coordination occur. In biological systems, these are signaling pathways or neural networks; in human-machine systems, they manifest as protocols, interfaces, and shared representational spaces \cite{Bradshaw2011}. These spaces enable ``joint activity'', characterized by inter-predictability, common ground, and directability.

\subsection{Integration Paradigms, Operational Dynamics, and Convergence Points}
Each paradigm embodies different principles from living systems theory:

In \textit{MAS}, collaboration occurs through well-defined protocols between autonomous agents, resembling distinct organisms in an ecosystem. Effectiveness emerges from collective behavior while maintaining individual identities.

In \textit{Centaurian systems}, integration blurs the boundaries between human and artificial components, forming functionally unified entities similar to biological organisms.

\newpage

The ecosystem versus organism analogy extends to operational dynamics:
\begin{itemize}
\item \textbf{Identity Maintenance}:
\textit{MAS} preserve distinct agent identities while allowing interaction while, 
\textit{Centaurian systems} create new composite identities through integration.

\item \textbf{Adaptation Mechanisms}:
\textit{MAS} adapt via reconfiguration of agent relationships, while
\textit{Centaurian systems} evolve through internal transformation of integrated components.

\item \textbf{System Boundaries}:
\textit{MAS} maintain clear boundaries with well-defined interfaces, while
\textit{Centaurian systems} develop permeable boundaries for deep functional integration.
\end{itemize}

Despite these distinctions, some scenarios blur the boundaries between para\-digms, especially in adaptive complex systems. Consider large action models (LAMs), where human operators provide training feedback to artificial systems. Depending on the context, these interactions can manifest as multi-agent collaboration or Centaurian integration. Such convergence points suggest that while the paradigms are conceptually distinct, they can complement each other in practice.

\subsection{Implications for System Design}
Viewing these paradigms through the lens of living systems theory has key design implications:
The \textit{MAS} design focus is on preserving agent autonomy, 
developing protocols for effective coordination, and ensuring system resilience through independent agents.
\textit{Centaurian systems}, on the other hand, have a design focus on emphasizing functional integration mechanisms, developing shared representational spaces, and encouraging emergent capabilities.

Section~\ref{sec:comm-spaces} will introduce communication spaces as a unifying framework supporting both paradigms, thus enabling both the loose coupling of MAS and the tight integration of Centaurian systems while preserving their core attributes.

\subsection{Situating Human-Machine Collaboration in the Context of Agentic AI}
Advances in agentic AI heighten the relevance of multi-agent and Centaurian systems. Unlike traditional AI chatbots, agentic AI systems engage in sophisticated reasoning and iterative planning to solve complex, multi-step problems autonomously. These systems follow a four-step cycle of \textit{perception}, \textit{reasoning}, \textit{action}, and \textit{learning}, continuously improving over time.

Our framework provides a theoretical foundation for integrating agentic AI into human-machine collaboration. The multi-agent paradigm aligns with contexts where agentic AI must maintain autonomy while coordinating with human and artificial agents via defined protocols---such as in distributed customer service or collaborative software development. 
Conversely, the Centaurian paradigm is crucial for cases requiring deep integration with human operators, such as healthcare or video analytics, where AI seamlessly merges with human expertise and decision-making. The tight coupling characteristic of Centaurian systems enables fluid interactions for complex tasks requiring human judgment alongside 
AI-powered analysis.

As agentic AI systems grow more sophisticated, the interplay between these paradigms becomes more dynamic. For instance, an enterprise AI system might operate independently (multi-agent mode) for routine tasks while tightly integrating with human experts (Centaurian mode) for complex decisions. Our framework supports the understanding and design of such interactions.

Additionally, the ``data flywheel'' effect---continuous learning and adaptation 
---aligns with our model's emphasis on system evolution. Learning occurs both in the improvement of independent agents (multi-agent paradigm) and in the refinement of human-AI integration patterns (Centaurian paradigm), demonstrating how our framework accommodates various aspects of agentic AI development.

%%%%%%%%%%%%%%%%%%%%%%%%%%%%%%%%%%%%%%%
\section{Technical Background: Petri nets}
\label{sec:tech-background}
%%%%%%%%%%%%%%%%%%%%%%%%%%%%%%%%%%%%%%%
To provide a rigorous foundation for modeling the complex interactions in both multi-agent and Centaurian systems, we employ Petri nets as our core formal framework. 
Petri nets present several advantages for our purpose: they offer 
a clear graphical representation while maintaining mathematical precision, they naturally capture concurrent processes and synchronization requirements, and they can be extended to handle sophisticated data types and conditions through their colored variants. These properties make them especially well-suited for modeling interactions among heterogeneous agents, whether in loosely coupled multi-agent configurations or tightly integrated Centaurian systems. This section introduces the key concepts of Petri nets and their extensions, which will serve as the formal basis for our communication spaces framework.

Petri nets are a well-established formalism for describing and analyzing the flow of information and control in concurrent systems. Originally introduced by \cite{Petri1962}, they have evolved into a family of models widely applied in computer science, systems engineering, and workflow management. A Petri net consists of \textit{places}, \textit{transitions}, and \textit{arcs} connecting places to transitions (or vice versa), with the system’s state captured by \textit{tokens} that reside in places. As transitions fire, tokens move among places, thus modeling the progress of a process or distributed computation \cite{Murata89}.

\subsection{Basic Concepts and Advantages}

\begin{itemize}
    \item \textbf{Graphical Clarity}:\ 
    One of the key strengths of Petri nets is their visual nature. Processes, resource constraints, and synchronization points are represented in a diagram, offering an intuitive view of concurrent interactions. This makes Petri nets well-suited for communicating system designs to both technical and non-technical stakeholders.

    \item \textbf{Explicit Concurrency}:\ 
    Petri nets explicitly capture concurrency by allowing multiple transitions to fire independently if they have sufficient tokens in their input places. This feature provides a natural way to represent multi-agent or multi-component systems where actions may proceed in parallel.

    \item \textbf{Formal Analysis Tools}:\ 
    A rich ecosystem of analysis techniques (e.g., reachability, liveness, boundedness) and software tools (e.g., colored Petri nets Tools\footnote{\url{https://cpnide.org/}}) exists for verifying properties of Petri net models \cite{Jensen1995}. This is particularly useful for systems requiring rigorous guarantees, such as safety-critical applications or complex human–AI workflows.

    \item \textbf{Modularity and Extensibility}:\ 
    Petri nets provide a modular framework that can be extended to various specialized forms (e.g., timed, stochastic, colored), allowing the designer to incorporate specific features such as timing constraints, probabilistic behavior, or typed tokens.
\end{itemize}

\subsection{Colored Petri Nets and Typed Tokens}
\label{subsec:colored-petri-nets}

While classical Petri nets use indistinguishable tokens to mark places, \textit{colored} Petri nets extend this model by assigning \textit{types} or \textit{colors} to tokens. Places become typed containers, and transitions can incorporate guard functions that enable more complex behaviors \cite{Jensen1995}. This extension allows designers to encode complex data structures and interaction protocols directly in the net, significantly increasing expressiveness. In the context of hybrid agent systems, colored Petri nets are especially advantageous for:

\begin{itemize}
    \item \textbf{Heterogeneous Agents}:\ 
    Different types of tokens can represent messages, tasks, or capabilities unique to human and synthetic agents.

    \item \textbf{State and Context Tracking}:\ 
    Color sets can capture the internal state of agents or system contexts, allowing transitions to fire only when certain conditions (guards) on those states are met.

    \item \textbf{Protocol Definition and Enforcement}:\ 
    Complex interaction sequences, including the dynamic creation and distribution of tasks, can be succinctly specified and analyzed.

    \item \textbf{Integration with Communication Spaces}:\ 
    As we will see in Section~\ref{sec:comm-spaces}, communication spaces---surface, observation, and computation---can be map\-ped onto distinct parts of a colored Petri net, leveraging types to separate or coordinate different communication and processing domains.
\end{itemize}

\subsection{Illustrative Example}

\begin{figure}[ht]
 \centerline{\includegraphics[width=0.8\textwidth]{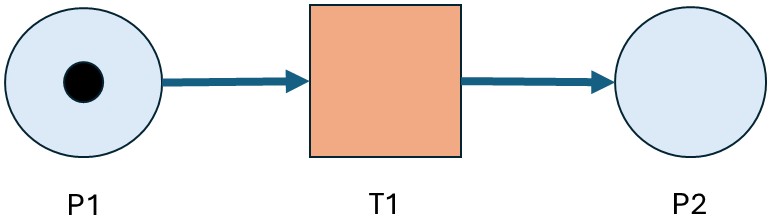}}
\caption{A simple Petri net example showing two places and one transition.}
    \label{Fig_Petri1}
\end{figure}

Figure~\ref{Fig_Petri1} presents a simplified Petri net modeling a small concurrent process. Places are shown as circles, transitions as rectangles, and arcs indicate the flow of tokens. 

\begin{figure}[ht]
 \centerline{\includegraphics[width=0.8\textwidth]{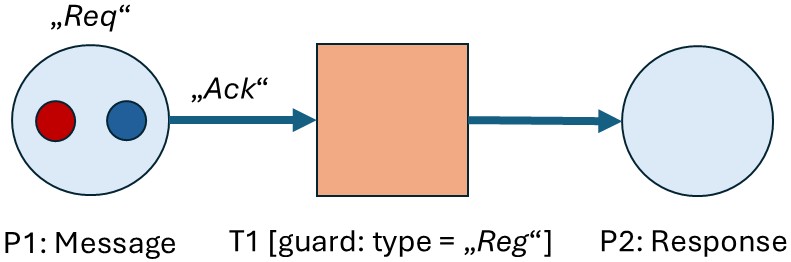}}
 \caption{A colored Petri net where tokens carry types or data, allowing conditional transitions.}
    \label{Fig_Petri2}
\end{figure}

In a colored Petri net version (Figure~\ref{Fig_Petri2}), tokens carry additional data types that reflect the roles or tasks assigned to different agents.

\subsection{Relevance to Hybrid Agent Systems}

Petri nets, and particularly their colored variants, are ideally suited to represent the parallel and event-driven nature of interactions among humans and AI components. Their token-based semantics naturally handles concurrency, while the ability to extend tokens with data types makes them powerful enough to capture the complexities of heterogeneous, hybrid agent systems. 

As we will see in the next section, Petri nets' properties---particularly their ability to model concurrent processes, handle heterogeneous data types, and support both loose and tight coupling---make them ideal for implementing the communication spaces framework, which addresses the coordination challenges identified in our discussion of multi-agent and Centaurian paradigms.

%%%%%%%%%%%%%%%%%%%%%%%%%%%%%%%%%%%%%%%
\section{Communication Spaces and Agent Architectures}
\label{sec:comm-spaces}
%%%%%%%%%%%%%%%%%%%%%%%%%%%%%%%%%%%%%%%
The contrasting requirements of multi-agent and Centaurian paradigms---main\-tain\-ing clear boundaries in one case while enabling deep integration in the other---call for a unified framework that can support both approaches. While Petri nets provide a strong formal foundation for modeling processes and synchronization, we need to extend their capabilities to capture the complexities of human-AI collaboration fully. We introduce \textit{communication spaces} as this unifying framework, building on Petri nets while adding specialized structures for handling varied communication styles, heterogeneous capabilities, and adaptive protocols.

\subsection{Foundations and Rationale}

Petri nets excel at modeling token flows among well-defined places, but in hybrid human-AI systems, we face additional challenges. Agents differ widely in perceiving, processing, and sharing information: humans communicate through natural language, AI systems through structured data, and physical devices through sensor signals. Moreover, these communication patterns must adapt dynamically as the system shifts between loose coupling (characteristic of multi-agent systems) and tight integration (needed in Centaurian configurations). These requirements call for a framework that can accommodate asymmetric interaction capabilities while supporting both autonomous operation and deep cognitive fusion.

\subsection*{A Mini-Example of Heterogeneous Agents}
\begin{figure}[ht]
 \centerline{\includegraphics[width=.9\textwidth]{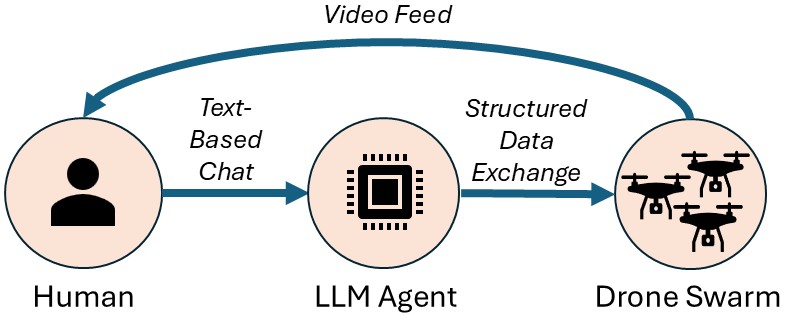}}
 \caption{Conceptual illustration of a human operator, an LLM agent, and a swarm of drones, each using distinct communication channels (text-based, structured data, and video feed).}
    \label{Fig_HumLLMDrone}
\end{figure}

Consider a team as shown in Figure~\ref{Fig_HumLLMDrone} with a human operator, an LLM-based conversational agent, and a swarm of simple robotic drones. The human and LLM exchange strategic decisions via text-based conversation, while the drones communicate real-time state updates in a lightweight sensor-data format. Additionally, the human observes a video feed of drone activity. Each of these ``channels'' imposes different synchronization rules, data formats, and reliability guarantees---precisely the nuances communication spaces help formalize.

\subsection{Core Ideas of Communication Spaces}

Communication spaces group interactions into three conceptual layers \cite{EurocastBBP24}: \textit{surface}, \textit{observation}, and \textit{computation}. Whether in a multi-agent (MAS) or a Centaurian system, each space encapsulates a coherent set of interaction rules and constraints.

The formation of these spaces typically involves either point-to-point or broadcast communication\footnote{Point-to-point or broadcast communication often uses message passing as a crucial concept in many system interactions, especially in distributed systems, concurrent programming, or event-driven architectures like ours. It allows different components (agents, large action models, etc.) to communicate without internal knowledge of the other system (decoupling), making it easy to add more components to a messaging-coupled system (scalability). Even different messaging protocols can be easily implemented and exchanged spontaneously (flexibility), with some protocols allowing asynchronous interaction in which the sender and receiver do not have to work simultaneously (asynchrony). \cite{Xiao2020} presents a comprehensive overview and analysis of the most advanced blockchain consensus protocols based on message passing.}, or communication through the environment\footnote{Shared memory architectures can form such an environmental common ground in the \cite{Bradshaw2011} sense.}.
Several examples illustrate this multi-agent organization of communication spaces:

\begin{itemize}
    \item In Mobile Ad-hoc NETworks (MANETs), mobile devices communicate with one another via wireless links without relying on an underlying infrastructure: each device acts as both an endpoint and a router forwarding messages to devices within radio range. 
    Software support for MANETs typically presents a coordination layer that sends specific messages to individual agents to guarantee the execution of cooperative processes while considering the need for connectivity continuance. 
    In \cite{BDHM06}, a formal model of the coordination mechanisms was given in terms of High-Level Petri nets.
    \item Management of transportation networks requires maintaining complete information on the whereabouts of vehicles and the state of roads to be mapped on the topology of the transportation network (see \cite{CGG18} for a concrete example).
    Then, the overall state of the transportation system is given by the information of which vehicle is at which node of the network, information which has to be maintained consistently for each vehicle.
    \item In the classical work by \cite{DBS06}, a computational model was derived from how ants communicate information on trails to reach food sources by marking them through pheromones recognized by other ants. The computational model was, therefore, based on associating weights to paths to a result and strengthening those associated with a successful one.
\end{itemize}

\begin{figure}[ht]
\centerline{\includegraphics[width=.9\textwidth]{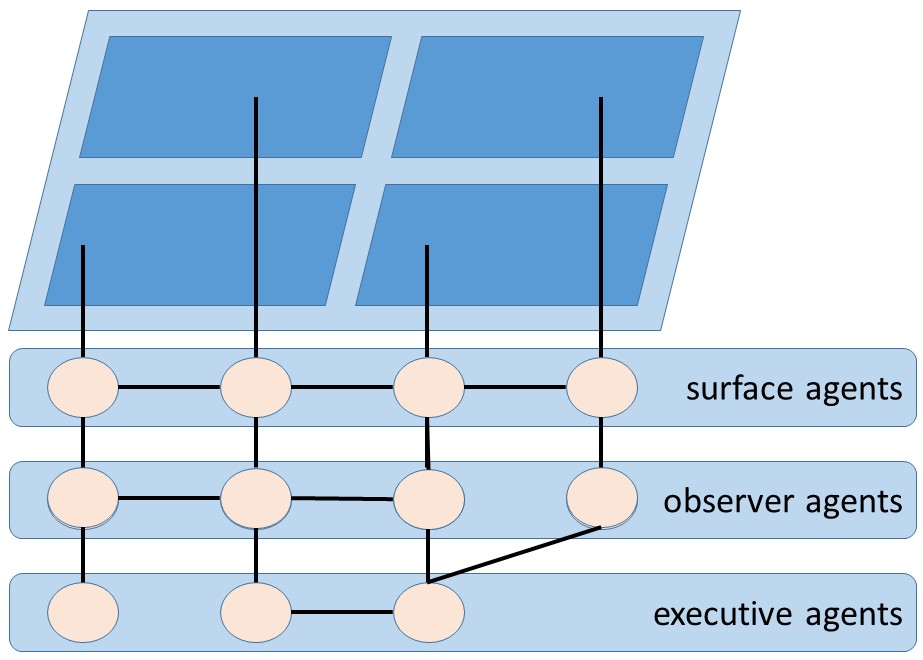}}
\caption{A layered architecture for HCI.} \label{Fig_LayeredArch}
\end{figure}

Figure~\ref{Fig_LayeredArch} shows a layered architecture similar to the {\em Model-View-Controller} (MVC) paradigm, where 
\begin{itemize}
\item {\em executive agents} implement the actual computations, 
\item {\em observer agents} realize a bridge between the computations, thereby managing the formatting of data for presentation and decoding user interactive commands and data entry, and 
\item {\em surface agents} (or mediators) are responsible for maintaining the presentation of the overall state to the users, thereby managing the materialization of the formatted data and the presentation layout, and providing support for capturing user interactions. 
\end{itemize}

Note that in this architecture, the actions performed by the executive agents, and not just their results, can be subject to observation.

\begin{figure}[ht]
\centerline{\includegraphics[width=\textwidth]{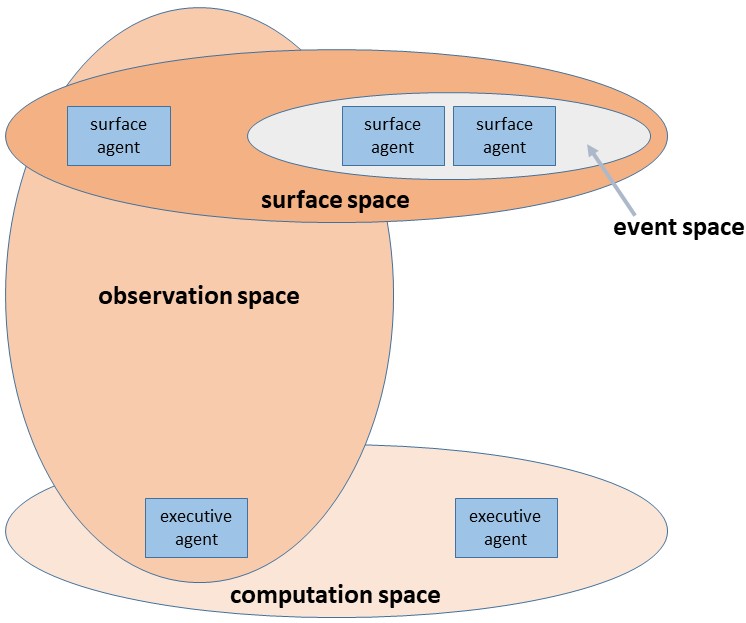}}
\caption{Schematization of communication spaces in the proposed architecture.} \label{Fig_HCI}
\end{figure}

Vertical structuring is also shown, relating surface agents and executive agents through an observation space (see Figure \ref{Fig_HCI}).

\subsubsection*{Surface Space}
This space mediates all contact with the outside environment---user interfaces, sensors, and external APIs. 

In MAS, surface space typically involves message-passing protocols or event listeners. 
In Centaurian systems, it can reflect a direct blending of human sensory input and AI-driven data capture.

\subsubsection*{Observation Space}
Bridging the surface interface with internal processing, observation space handles message transformations, routing, and light coordination. 

In MAS, protocols here ensure agents remain autonomous yet cooperative. 
In Centaurian systems, observation may feature continuous feedback loops that unify human perception with AI analysis.

\subsubsection*{Computation Space}
Serving as the system’s “core,” the computation space performs decision-making, allocates resources, and generates final outputs. 

MAS solutions often involve multiple autonomous modules, each coordinating a portion of the computation. 
By contrast, Centaurian architectures might fuse human insights with AI algorithms in a shared decision-making environment.

\newpage

\subsubsection*{Flows and Feedback}
Information travels \textit{vertically} through these spaces---from raw input at the surface to internal processing and back. There may also be \textit{horizontal} flows where agents collaborate within a given space. Feedback loops enable adaptive behavior, such as refining protocols or redistributing tasks in real time based on performance.

\begin{figure}[ht]
\centerline{\includegraphics[width=.85\textwidth]{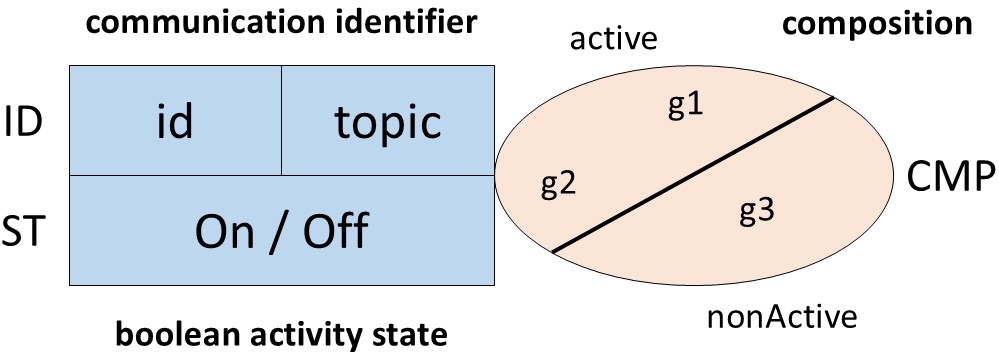}}
\caption{A schematization of the basic structure of a group agent.}\label{Fig_GroupAgent}
\end{figure}

Each of these spaces can be seen as managed by an associated \textit{group-agent}, facilitating distributed problem-solving among a group of users within a networked system or application.
The structure of a group-agent is schematized in Figure~\ref{Fig_GroupAgent}, where the \texttt{ID} and \texttt{ST} compartment are common to all agents.
Then, the \texttt{COMPOSITION} compartment is specific to a group-agent, while other agents have specific additional members for defining their state, and are endowed with a set of behaviors.

Group-agents are relevant in collaborative systems where multiple users and agents work together to solve complex problems or tasks and proper delivery of messages is crucial to ensure that each participant is correctly informed of the overall progress towards task achievement.
A group-agent can be created at any time with respect to a topic and its \texttt{register} behavior regulates composition modification through a protocol which respects the localization constraints.
The following pseudo-code provides an abstract specification of the construction of group-agents for communication spaces.
\medskip

\noindent
{\tt\small
register(self, {\bf agent}) \{  \newline 
\hspace*{1cm}
	if {\bf agent} is concerned with self.\textbf{topic} then  \newline
		\hspace*{2cm} if {\bf agent}.ST = ON then \newline 
	    \hspace*{3cm} self.CMP[active].Add({\bf agent}) \newline
	    \hspace*{2cm} else self.CMP[nonActive].Add({\bf agent}) \newline
\hspace*{1cm} \} \newline
\}   
}
\medskip

Once created, a group-agent has to guarantee that any message from a member of the group can be read by all other (active) members.
In particular, a message sent by an agent registered to a group and concerning the group's topic is delivered to all active agents registered to that group, as for the group's \texttt{deliver} behavior, to be invoked on message generation.  
\smallskip

\noindent
{\tt\small
deliver(m}) \{  \newline 
\hspace*{1cm}
if self.ST == ON and self.isConcernedWith(m.\textbf{topic}) then \newline
    \hspace*{2cm} for each \textbf{agent} in self.CMP[active] \newline
    \hspace*{3cm} \textbf{agent}.addToBuffer(m) \newline
\hspace*{1cm} \newline
\}   
\medskip

Similarly, deregistration can occur, when a component is no longer relevant to a group topic, or due to an explicit decision of the coordination mechanism.
Deregistration can only occur for active agents.
\smallskip

\noindent
{\tt\small
deregister(self, {\bf agent}) \{  \newline 
\hspace*{1cm}    if {\bf agent} $\in$ self.CMP[active] then \newline
\hspace*{2cm}    self.CMP[active].Remove({\bf agent})
\newline
\}   
}
\medskip

The information about the activity status of an agent registered to a group is updated through the \texttt{switch} behavior

\smallskip

\noindent
{\tt\small
switchCMP(self, {\bf agent}) \{ \newline
\hspace*{1cm}    if {\bf agent}.ST == OFF and {\bf agent} $\in$ self.CMP[active] then \newline
\hspace*{2cm}    self.CMP[active].Remove({\bf agent}) \newline
\hspace*{2cm}    self.CMP[nonactive].Add({\bf agent}) \newline
\hspace*{1cm}    if {\bf agent}.ST == ON and {\bf agent} $\in$ self.CMP[nonactive] then  \newline
\hspace*{2cm}    self.CMP[nonactive].Remove({\bf agent}) \newline
\hspace*{2cm}    self.CMP[active].Add({\bf agent})
\hspace*{1cm} \newline
\}
}

\subsection{Implementing Communication Spaces}

Each of the three spaces can be mapped onto subsets of Petri net places (or subnets) with typed tokens indicating relevant data, tasks, or events. This approach partitions a net into self-contained ``zones'', each governed by specific interaction rules and constraints.

\subsubsection*{Multi-Agent Architecture}
In a multi-agent system, \textit{surface agents} manage user interactions and detect events from the environment, acting as the system's interface with external inputs. \textit{Observer agents} ensure smooth communication by transforming and routing messages, enabling coherent collaboration among autonomous components. Meanwhile, \textit{executive agents} operate within the computational space, executing specialized tasks and maintaining a partial system state to support overall functionality.

A key benefit is clear boundary maintenance and scalability: each agent type can be added or removed with minimal impact on the others, provided communication protocols remain consistent.

\subsubsection*{Centaurian Architecture}
In Centaurian systems, a \textit{unified interface layer} merges human and AI sensory input within the surface space, creating a seamless interaction point. A \textit{shared observation layer} acts as a bridge between human and AI representations, facilitating smooth cognitive fusion. Meanwhile, a \textit{hybrid processing layer} integrates human reasoning with algorithmic methods, allowing for emergent collective problem-solving.

Such architectures thrive on adaptability, where boundaries are permeable and agents can fluidly exchange roles or data to optimize performance.

\subsection{Formal Representation with Colored Petri Nets: Communication Space Petri Net}
We extend colored Petri nets to create a \textit{Communication Space Petri net}, where places are partitioned into \textit{surface}, \textit{observation}, and \textit{computation} categories.
Tokens carry types or ``colors'' corresponding to data formats or message semantics relevant to each space.
Transition guards incorporate rules for agent-specific capabilities (e.g., a color set may denote ``human text command'' vs. ``drone sensor data'').

This enriched formalism preserves the tractability of standard Petri nets while accommodating the specialized interactions of heterogeneous agents. For example, transitions in the observation space can fire only when tokens of type ``parsed command'' are present, ensuring that raw data never bypasses the necessary transformation steps.

\subsubsection*{Example Snippet}
\begin{figure}[ht]
\centerline{\includegraphics[width=0.8\textwidth]{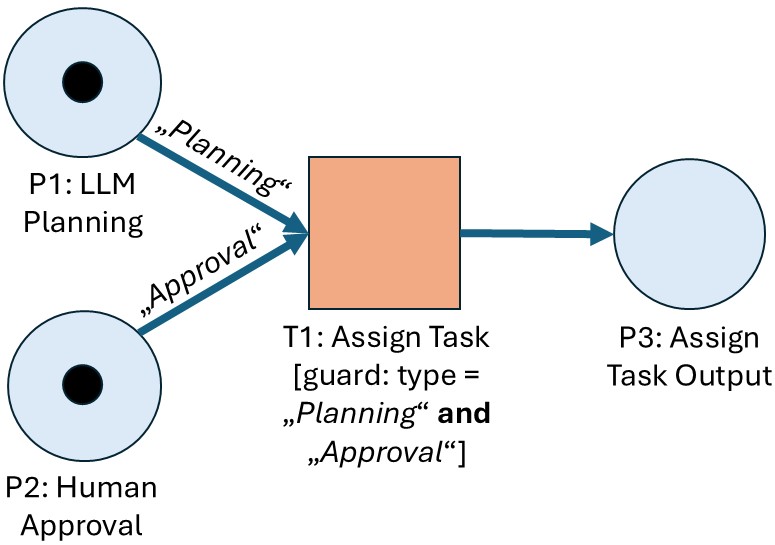}} 
    \caption{A schematic colored Petri net snippet illustrating how a transition fires only if both an LLM planning token and a human approval token are present, capturing Centaurian synergy.}
    \label{Fig_Petri3}
\end{figure}

Imagine a transition as shown in Figure~\ref{Fig_Petri3} in the computation space labeled \textit{Assign Task}, which fires only when it receives both a ``planning token'' from the LLM agent and an ``approval token'' from the human. Such a rule encapsulates the synergy sought in a Centaurian design, formalizing how human sign-off triggers an AI-based planning process.

\subsection{Implementation Considerations}

Implementing communication spaces requires careful attention to the following aspects:
\textit{State management} ensures consistent updates across distributed agents or fused human–AI modules, especially in time-sensitive environments.  
\textit{Resource allocation} deals with balancing computational load, bandwidth, and potential real-time constraints.  
\textit{Protocol adaptation} supports switching between agent autonomy (MAS) and deeper integration (Centaurian) based on changing system goals or unexpected conditions.

A system may operate predominantly in MAS mode for routine tasks, yet temporarily shift to a more Centaurian-like strategy when it encounters novel or ambiguous conditions. This capacity to blend paradigms on demand underscores the flexibility of communication spaces.

The benefits can be summarized as follows:
By demarcating surface, observation, and computation spaces within a unified Petri net model, designers gain a clear blueprint for structuring agent interactions. This approach enables them to capitalize on the strengths of both multi-agent and Centaurian paradigms---maintaining well-defined boundaries when necessary, but also allowing for deep integration and emergent capabilities where advantageous. The next sections illustrate how communication spaces operate in real-world scenarios, further emphasizing their role in building robust human-AI systems.

%%%%%%%%%%%%%%%%%%%%%%%%%%%%%%%%%%%%%%%
\section{Use Case 1: Multi-agent Interaction with Satellite and Swarm Robots}\label{sec:usecase1}
\begin{figure}[ht]
\centerline{\includegraphics[width=.8\textwidth]{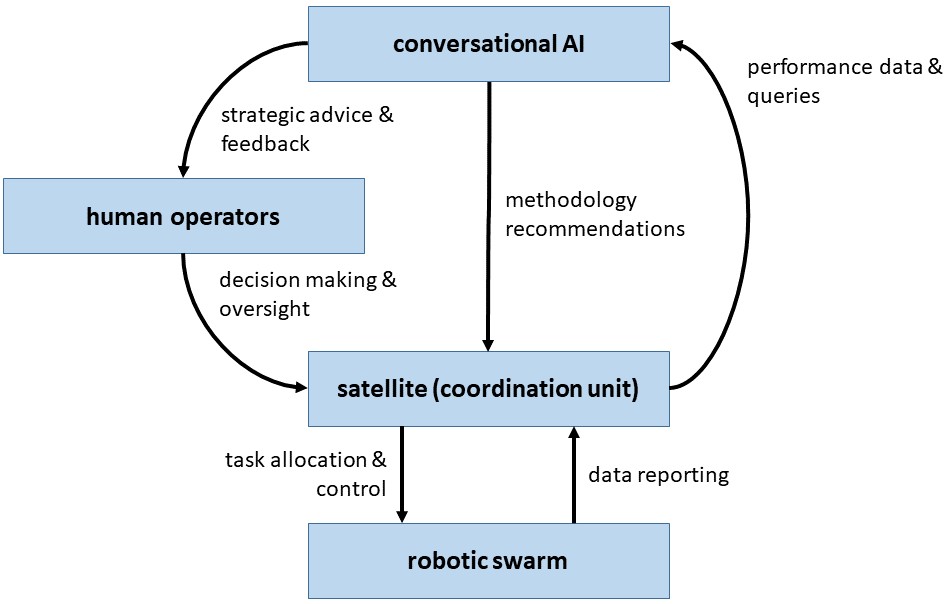}}
\caption{Interaction of human operators, conversational AI, satellite, and a robotic swarm.} \label{Fig_Swarm}
\end{figure}

%%%%%%%%%%%%%%%%%%%%%%%%%%%%%%%%%%%%%%%
This use case demonstrates how our theoretical framework accommodates both multi-agent and Centaurian paradigms within a complex system. While predominantly exhibiting multi-agent characteristics through its distributed architecture, the system also incorporates Centaurian elements in specific human-AI interaction points. Figure~\ref{Fig_Swarm} illustrates the data flow in an experiment with a semi-centralized coordinated swarm of robots, using both ``rigid'' optimization algorithms and ``flexible'' intervention through a large language model (LLM). This setup encapsulates a true multi-agent HCI interaction, integrating human operators, conversational AI, the satellite control unit, and swarm robots.

This use case extends the application of a semi-centralized control strategy, originally enhanced with blockchain technology for security and transparency \cite{carovilla2023integrating}. The multi-agent paradigm is evident in how distinct entities---the satellite control unit, swarm robots, LLM, and human operators---maintain their functional independence while collaborating through well-defined protocols. However, the system also exhibits Centaurian characteristics in specific interaction points, particularly in the human-LLM interface. 

\begin{figure}[ht]
\centerline{\includegraphics[width=\textwidth]{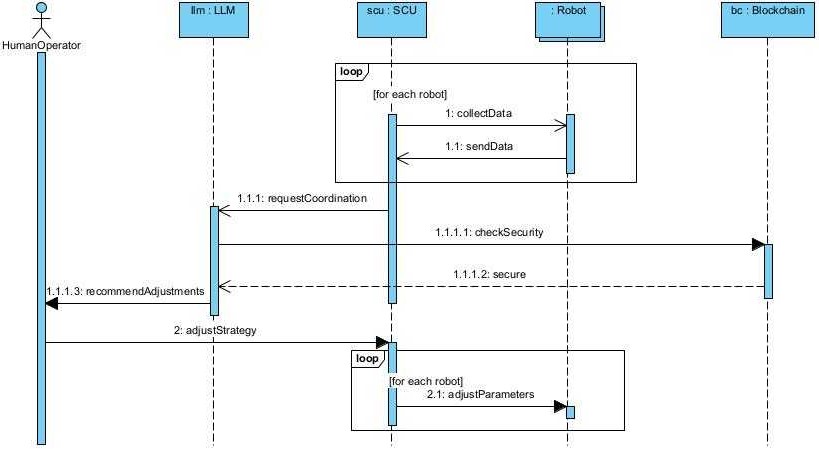}}
\caption{Information and decision flows in the semi-centralized control strategy.}
\label{Fig_Swarm2}
\end{figure}

The overall information and decision-making flow, described in detail in the following sections, is shown in Figure~\ref{Fig_Swarm2}.

\subsection{System Overview}
The system exemplifies the multi-agent paradigm through its composition of distinct, autonomous components: a satellite control unit, a swarm of robotic agents, an LLM, and human operators. 

Each agent maintains its own decision-making capabilities while participating in the collective process. The satellite provides centralized coordination and processes the data collected by the swarm, while the LLM offers strategic advice on algorithm selection and parameter tuning. Human operators, rather than being mere supervisors, function as autonomous agents who monitor these interactions and intervene to guide the system through complex decision-making processes.

\subsection{Integration with LLM and Human Operators}
The integration layer demonstrates a hybrid approach combining both paradigms. Following the multi-agent paradigm, the LLM interacts with the blockchain and control algorithms as an independent agent, providing a flexible intelligence layer. However, the interaction between human operators and the LLM exhibits Centaurian characteristics, as it creates a tightly coupled decision-making unit where human judgment and AI recommendations are seamlessly integrated. 

This dual-nature system remains robust and adaptive through both the independent operation of its components and the synergistic human-AI integration points.

\subsection{Practical Applications}
The practical applications highlight how the two paradigms complement each other. In path optimization and adaptive problem solving, the system primarily operates in a multi-agent mode, with the LLM functioning as an independent advisor. However, when human operators review and implement suggestions, the interaction shifts temporarily to a Centaurian mode, creating a tightly integrated human-AI decision-making unit. 

For example, in path planning, the LLM can suggest changes to curve parameters based on detected obstacles or changes in terrain, maintaining its agent independence, while human operators review these suggestions through a Centaurian-style deep integration process.

\subsection{Human-Agent Collaboration}
The collaboration model in this system demonstrates the value of supporting both paradigms. While the overall architecture follows multi-agent principles, with human operators functioning as independent agents within the system, their interaction with the LLM and control systems often exhibits Centaurian characteristics. This flexible approach allows the system to leverage both the independence and coordination capabilities of the multi-agent paradigm and the deep integration benefits of the Centaurian approach. The result is a robust system that protects against the risks of automation while actively leveraging the unique strengths of both humans and machines, pushing the boundaries of what technology can achieve through this dual-paradigm approach.

%%%%%%%%%%%%%%%%%%%%%%%%%%%%%%%%%%%%%%%
\section{Use Case 2: Large Action Models (LAMs) through Feedback Loops on HCI Interactions}\label{sec:usecase2}
%%%%%%%%%%%%%%%%%%%%%%%%%%%%%%%%%%%%%%%
While our first use case emphasized the multi-agent paradigm, this second use case demonstrates a stronger inclination toward the Centaurian approach, while still maintaining some multi-agent characteristics. 

\begin{figure}[ht]
\centerline{\includegraphics[width=.8\textwidth]{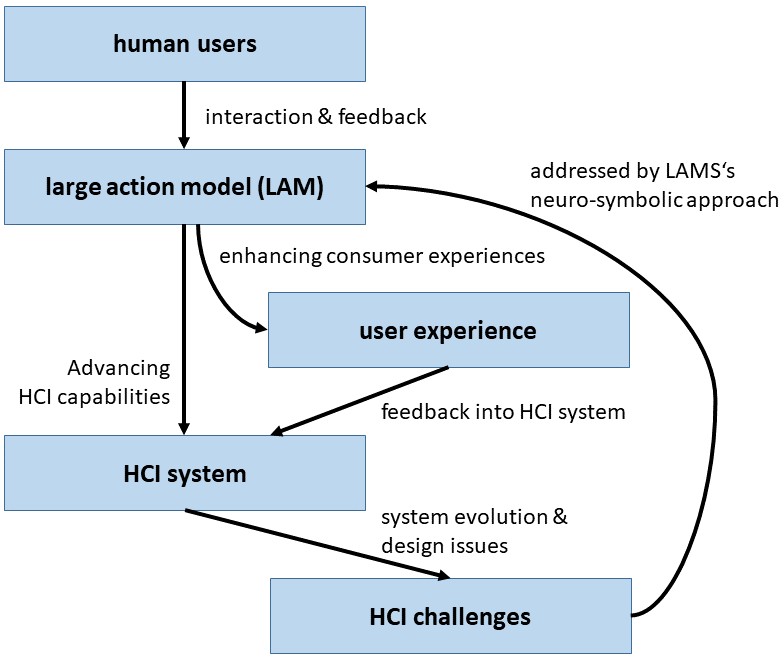}}
\caption{Feedback loop for improved agent interactions.}\label{Fig_LAM}
\end{figure}

Figure~\ref{Fig_LAM} shows the framework of {\sc Rabbit Tech}'s Large Action Model (LAM)\footnote{\url{https://www.rabbit.tech/research}}, which integrates advanced computational agents to effectively model and predict human actions on computer applications. The system exemplifies the Centaurian paradigm's emphasis on tight integration between human and artificial components, while incorporating multi-agent elements in its distributed architecture.

This use case illustrates the transformative potential of LAMs in creating a new paradigm for human-computer interaction. Unlike traditional multi-agent systems where components maintain clear boundaries, Rabbit LAM demonstrates the Centaurian ideal of deep functional integration, continuously learning from user input and refining its predictive capabilities. It embodies a sophisticated feedback loop system where the boundaries between human and artificial intelligence become increasingly fluid, reflecting the core principle of Centaurian systems where components form an integrated whole rather than remaining distinct entities.

\subsection{System Overview}
The LAM framework at {\sc Rabbit} API exemplifies the Centaurian paradigm through its core components: the LAM node, human users, and the HCI system, together with representations of the challenges this integration addresses and of the user experience. Rather than maintaining strict boundaries between these components, they evolve together through their interactions. The LAM node, while functioning as a computational agent, achieves a level of integration with human users that transcends traditional multi-agent relationships, using neuro-symbolic programming to create a unified decision-making entity.

\subsection{Neuro-Symbolic Integration}
The neuro-symbolic approach of LAM represents a key Centaurian characteristic: the seamless integration of different processing paradigms. By combining neural network capabilities for pattern recognition with symbolic AI for rule-based processing, LAM demonstrates how apparently distinct computational approaches can be unified into a coherent whole, much like the human-AI integration central to Centaurian systems. This integration enables complex task execution across multiple applications with high accuracy and minimal latency.

\subsection{Human-Agent Interaction}
The human-LAM interaction epitomizes the Centaurian ideal of symbiotic integration. Through a natural language-based user interface that processes spoken or typed commands, the system achieves a level of interaction where the boundaries between human intent and machine execution become increasingly blurred. The feedback system allows LAM to learn from each interaction, adapting its models to better match user preferences---a hallmark of Centaurian systems where components evolve together rather than merely coordinating their actions.

\subsection{Practical Applications and Challenges}
LAM's approach to task management and problem solving demonstrates the advantages of Centaurian integration over purely multi-agent coordination. By anticipating user needs and providing proactive solutions, it shows how deep human-AI integration can surpass the capabilities of systems where components merely coordinate their actions. However, this tight integration brings unique challenges, including ensuring AI decision transparency and maintaining privacy---challenges characteristic of Centaurian systems where the boundaries between human and artificial components become less distinct.

\subsection{Key Takeaways}
The feedback loops in LAM exemplify the evolutionary nature of Centaurian systems. Unlike multi-agent systems where components maintain fixed identities and interact through stable protocols, LAM's feedback loops create a dynamic system where human and artificial components grow together, continuously refining their interaction patterns. This approach to HCI demonstrates how Centaurian principles can create more intuitive and responsive computing environments, while still benefiting from certain multi-agent characteristics in its distributed architecture.

%%%%%%%%%%%%%%%%%%%%%%%%%%%%%%%%%%%%%%%
\section{Related Work}\label{sec:relwork}
%%%%%%%%%%%%%%%%%%%%%%%%%%%%%%%%%%%%%%%
The implementation of agent interactions has been explored through various computational approaches. Basic interaction protocols have been formalized for distributed computation \cite{AspnesR09}, while more complex patterns have been developed for knowledge combination \cite{AndreoliBP94,BorghoffS96} and distributed problem-solving \cite{BorghoffPAF98}. Recent work has focused on higher-level interactions, particularly in intention recognition \cite{GreefDGL07,IngaRRNRKDLTNMH23}, which is becoming increasingly important as artificial agents grow more sophisticated in their ability to understand and respond to human goals.

\cite{GuamanDP21} presents an innovative unsupervised clustering theory using self-or\-ga\-niz\-ing maps to classify \textit{MVC} patterns based on software quality metrics in a continuous interaction process. Their goal is to identify quality features that determine the similarity of MVC applications without architectural bias.
\cite{LawlessS21} use interdependence theory to determine whether convergence has a positive or negative effect in a competition between human and artificial agents, as in system dynamics models. 

Mobile applications, as well as satellite and swarm robotics, have grown significantly in recent years, with architectural patterns playing a critical role in their success. To ensure the correct implementation of an MVC pattern, \cite{DobreanD22} propose an automated technique that analyzes and detects architectural issues using data from {\em Software Development Kits} (SDKs) specifically targeted at mobile code bases.
To enhance coordination and security in swarm robotic systems, \cite{carovilla2023integrating} introduce a semi-centralized framework that integrates blockchain technology, featuring a centralized control unit that coordinates the swarm, while blockchain technology ensures secure and decentralized data storage and communication.
\cite{KumarC23} take the opposite approach, extending pattern descriptions with a {\em System of Systems} (SoS) model to apply interactive dynamics to an MVC pattern.

Using the {\em Internet of Things} (IoT) paradigm as a complex distributed system that shares a 3-tier architecture consisting of embedded nodes, gateways that connect an embedded network to the wider internet, and data services in servers or the cloud, \cite{RiliskisHL15} propose a novel approach for programming applications across 3-tiers using a distributed extension of the MVC architecture. 
Also in the IoT context, with a focus on the gaming/sports arena, \cite{Chen25} sees 
%
%great 
%
potential for entertainment robots to recognize human posture through artificial intelligence. 
By detecting and monitoring users' movements in real time, these robots provide a personalized and interactive entertainment experience.
\cite{CaciD20} also discuss the interplay between games with personality and artificial intelligence. They use the term ``virtual human'' to describe a computer program that simulates a human in some aspects.

With the ubiquity of AI applications, HCI research is increasingly integrating these approaches, demonstrating that AI and HCI are mutually beneficial when they collaborate \cite{PandaR24}.
However, criticism and bad experiences have also been reported. 
For example, \cite{Choudhuri2024} state that there were no statistical differences in participants' productivity or self-efficacy when using ChatGPT compared to traditional resources.
Instead, they found significantly higher frustration levels, identifying five distinct errors resulting from violations of human-AI interaction guidelines, leading to various (negative) consequences for the participants.

LLMs are increasingly being used in studies to investigate the interaction between humans and artificial agents. 
\cite{BorghoffMM24} explore human-artificial interaction with generative AIs in a software engineering project course.
In the same domain, \cite{Nascimento2023} evaluates ChatGPT-generated code against developer-generated code to determine which tasks are better suited for engineers and which are better handled by AI. This could lead to more efficient interaction (e.g., AI as a tutor for SE developers) and provide new insights into innovative AI strategies that include the involvement of {\em humans-in-the-loop} to support the tasks of software engineering.
\cite{CapitanelliM24} demonstrate that LLMs can play an important role in planning actions in human-robot interactions.

\cite{GiudiciLTB24} present and evaluate a method for analyzing user reactions to AI using a live-streaming platform where human streamers conduct interviews that are transmitted to a specially developed GPT voice interface using a crowd-based approach.
\cite{KimI23} study the attribution of human characteristics to artificial intelligence. They developed a tool to measure how users form anthropomorphic reactions to interactions with AI chatbots in a banking service setting. 
For a survey of {\em Artificial Emotional Intelligence} (AEI) for cooperative social human-machine interactions see \cite{AhmadiH23}.

Message Passing Neural Networks (MPNNs) \cite{papillon2023} are a special class of neural networks that use the aforementioned message-passing paradigm to capture interactions between entities (often represented as nodes in a graph).
In MPNNs, nodes in a graph communicate with their neighbors through a series of message-passing steps. This interaction allows the network to learn representations by aggregating information from neighboring nodes; see also the \textsc{Rabbit} use case in Section~\ref{sec:usecase2}.

A formal model of the global activity of a system of agents is provided by \textit{concurrent game structures}, describing a situation where agents act according to individual \textit{strategies} by which to select individual actions based on knowledge of the global state of the system.
Reasoning on the possible evolutions of such systems leads to the definition of some variations of \textit{Alternating-Time Temporal Logic} (see e.g., \cite{AHK02,ALNR17,MMPV16}). In contrast, a logic that incorporates spatial constraints on a par with temporal ones has been recently proposed by \cite{BLP24}.
In general, these models abstract the actual mechanisms through which information about states can be exchanged among agents and assume that all agents act rationally regarding their strategies. 
The level of non-determinism inherent to considering human agents would, therefore, require some adaptation of these logics in the direction of identifying classes of agents that can follow several strategies at once.

Although beyond the scope of this paper, there exists a humanistic perspective \cite{KnellR24}, as well as an ethical dimension, to 
human-artificial interaction. 
For the latter, the reader is referred to \cite{KumarCx23,Karlan23} and \cite{Nieto24}.
In some ways, ``virtual superhuman''  AI technologies could undermine the value of human achievements \cite{SchaapSS24} or even may pose an existential threat \cite{KumarC23a}.

%%%%%%%%%%%%%%%%%%%%%%%%%%%%%%%%%%%%%%%
\section{Conclusion and Future Work}\label{sec:concls}
%%%%%%%%%%%%%%%%%%%%%%%%%%%%%%%%%%%%%%%
We have outlined a framework that systematically considers how human and artificial agents interact within a networked ecosystem. By articulating HCI not just as human-computer interactions but as dynamic interactions among different agents within a flexible, multi-agent system, we have highlighted the potential for more complex and effective interactions.
The application of our multi-agent model to two different use cases---satellite and swarm robotics, and Large Action Models (LAMs)---has not only validated our theoretical constructs, but also underscored the practical implications of such interactions. 

In the first use case, we observed how a semi-centralized system, enabled by a large language model, could improve decision-making processes and adaptability through real-time data-driven adjustments. This setup not only ensures efficiency, but also preserves human oversight, which is critical for managing complex, dynamic environments.

The second use case involving LAMs further highlighted the potential of integrating advanced computational agents to predict and shape human-computer interactions. Using neuro-symbolic programming, the LAM framework demonstrated its ability to refine its operations based on continuous feedback from human interactions, promoting an adaptive and responsive HCI system.

We have thus established a foundational framework for the design and analysis of HCI systems that views these interactions as a complex interplay among multiple agents. This approach allows us to better understand and harness the collective capabilities of different agents, thereby enhancing the system's adaptability and resilience. Importantly, this perspective aligns with emerging technology paradigms that foster a deep understanding and integration of human agents.
Indeed, our contribution can be seen as a systemic evolution in design approaches, emphasizing the front-end to enhance usability and effectiveness. This contrasts with, but complements, systemic approaches that prioritize the back-end for security and robustness, as illustrated by \cite{SalzanoPMT23}. Together, these perspectives synergistically improve the overall architecture of HCI systems.
Ultimately, our study enriches the understanding of HCI as a dynamic and evolving field characterized by the complex yet symbiotic relationships between humans and machines. As we delve deeper into these interactions, the possibility of transforming human-computer interactions into more seamless and effective experiences becomes increasingly tangible. This promises a future in which the human and artificial dimensions are not merely aligned but fully integrated in a balanced and complementary manner. Such integration could pave the way for a new era of collective and hybrid intelligence---where humans and machines work closely together---potentially the most significant and immediate outcome of current rapid technological advances.

Future research will focus on improving the adaptability of MAS by addressing the challenge of dynamic team composition. Standard Petri nets, with their static structure, have difficulty accommodating scenarios where agents join or leave while maintaining consensus, as illustrated by the \texttt{register} and \texttt{deregister} (for active agents) pseudo-code in Section~\ref{sec:comm-spaces}.
Therefore, we plan to model agents as individual (colored) Petri nets with specialized input and output locations, allowing token-based message passing for seamless registration and deregistration. However, smooth coordination requires high-level reconfiguration mechanisms to dynamically manage agent lifecycles. While these improvements enhance the adaptability of MAS, Centaurian systems, with their stable human-AI integration, remain well-suited to Petri net formalization due to their reliance on fixed coordination structures.

Another important future research direction is the development of a hybrid approach that bridges MAS and Centaurian systems. While Petri nets effectively model structured Centaurian intelligence, where human and AI components collaborate through predefined coordination rules, they fall short in dynamic MAS scenarios that require self-organization. Future research will therefore integrate high-level reconfigurable networks to enable agent fluidity, while using communication spaces as an intermediate layer for seamless coordination between autonomous and tightly coupled agents. This three-tiered approach will enable dynamic transitions between structured Centaurian intelligence and decentralized MAS interactions. Given the increasing importance of Centaurian AI in human-computer collaboration, this hybrid framework will improve efficiency, adaptability, and interoperability.

An alternative set-theoretic approach abstracts agent relationships without rigid network structures, defining interactions through set membership and relational mappings. While this circumvents the limitations of Petri nets in dynamic scenarios, its implementation requires structured protocols for message flow and consistency. Our future research will therefore explore how \textit{set-theoretic models} can complement Petri nets to ensure both formal rigor and practical applicability in human-AI collaboration.

%%%%%%%%%%%%%%%%%%%%%%%%%%%%%%%%%%%%%%%
%%%%%%%%%%%%%%%%%%%%%%%%%%%%%%%%%%%%%%%
\section*{Conflict of Interest Statement}
The authors declare that the research was conducted in the absence of any commercial or financial relationships that could be construed as a potential conflict of interest.

\section*{Author Contributions}
All authors listed have made equal, substantial, direct, and intellectual contributions to the work and have approved it for publication.

\section*{Funding}
Remo Pareschi has been funded by the European Union—NextGenerationEU under the Italian Ministry of University and Research (MUR) National Innovation Ecosystem grant ECS00000041-VITALITY—CUP E13C22001060006.

% \section*{Acknowledgments}
%

% \section*{Supplemental Data}
% \href{http://home.frontiersin.org/about/author-guidelines#SupplementaryMaterial}{Supplementary Material} should be uploaded separately on submission, if there are Supplementary Figures, please include the caption in the same file as the figure. LaTeX Supplementary Material templates can be found in the Frontiers LaTeX folder.

% \section*{Data Availability Statement}
% The datasets [GENERATED/ANALYZED] for this study can be found in the [NAME OF REPOSITORY] [LINK].
% Please see the availability of data guidelines for more information, at https://www.frontiersin.org/about/author-guidelines#AvailabilityofData



\end{document}